\documentclass[11pt, article, nofootinbib]{revtex4}
\usepackage{graphicx}
\usepackage{tcolorbox}
\usepackage{hyperref}
\usepackage{amsmath,amsthm, amssymb}

\newtheorem*{prob}{Problem}
\newtheorem{thm}{Theorem}
\newtheorem{corollary}{Corollary}
\newtheorem{lemma}{Lemma}

\begin{document}
\title{Dynamic Sampling from a Discrete Probability Distribution with a Known Distribution of Rates}
%
%\titlerunning{Dynamic Sampling from a Discrete Probability Distribution}
\author{Federico D'Ambrosio$^1$, 
Hans L. Bodlaender$^1$ and 
Gerard T. Barkema$^1$}
%
%\authorrunning{F. D'Ambrosio et al.}
\address{$^1$ Department of Information and Computing Science, Utrecht University, Princetonplein 5, 3584 CC Utrecht, The Netherlands
}
\begin{abstract}
In this paper, we consider several efficient data structures for the problem of sampling from a dynamically changing discrete probability distribution, where some prior information is known on the distribution of the rates, in particular the maximum and minimum rate, and where the number of possible outcomes N is large.

We consider three basic data structures, the Acceptance-Rejection method, the Complete Binary Tree and the Alias method. These can be used as building blocks in a multi-level data structure, where at each of the levels, one of the basic data structures can be used, with the top level selecting a group of events, and the bottom level selecting an element from 
a group.

Depending on assumptions on the distribution of the rates of outcomes, different combinations of the basic structures can be used. We prove that for particular data structures the expected time of sampling and update is constant when the rate distribution follows certain conditions. 
We show that for any distribution, combining a tree structure with the Acceptance-Rejection method, we have an expected time of sampling and update of $O\left(\log\log{r_{max}}/{r_{min}}\right)$ is possible, where $r_{max}$ is the maximum rate and $r_{min}$ the minimum rate. We also discuss an implementation of a Two Levels Acceptance-Rejection data structure, that allows expected constant time for sampling, and amortized constant time for updates, assuming that $r_{max}$ and $r_{min}$ are known and the number of events is sufficiently large.

We also present an experimental verification, highlighting the limits given by the constraints of a real-life setting.

\end{abstract}

\maketitle       % typeset the header of the contribution
\section{Introduction}
\begin{figure}[!hbt]
\begin{center}
 \includegraphics[width=.5\textwidth]{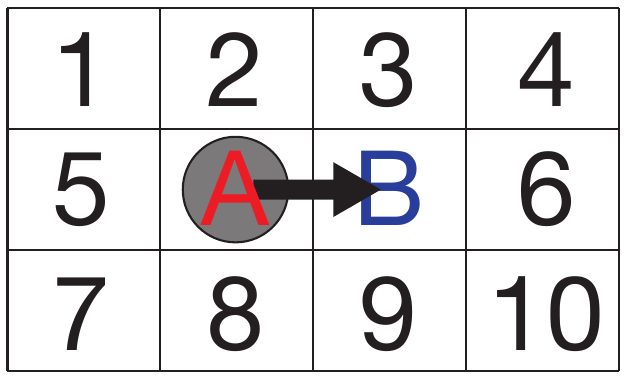}
 %\Description{An atom moves from position A to position B, surrounded by the ten neighboring sites, whose occupational state will determine the hopping rate for that move} 
 \caption{In the continuous time simulation of MBE growth on a metallic substrate, the hopping rate of an atom of copper from a position A to a position B is dependent on the occupational state of the ten surrounding sites: each move will influence the hopping rates of the surrounding atoms. It is critical to store these rates in a data structure that allows for updates. }
 \label{fig:atom}
 \end{center}
\end{figure}

\paragraph*{The problem}
In this paper, we consider the design of data structures for the following problem. We have a dynamic
discrete probability distribution, more precisely, we have a finite set of events, each with a
rate. We have the following operations on the data structure: an event can be deleted, inserted,
change its rate, and we want to randomly select an event, with each event selected with a 
probability proportional to its rate. This problem is well understood when the number of events
is small, but in many applications, we need to sample from a very large collection of events.

In this paper, we make one further assumption: we assume that the rates of possible events are distributed according to a known and unchanging probability distribution $\rho(r)$, i.e. the expected number of events with a rate between $r$ and $r+\Delta r$ out of $N$ total events can be computed as:
\begin{equation} 
E\left[ n_{r,r+\Delta r}\right] = N \int_r^{r+\Delta r} \rho(x) \, dx.
\end{equation}

From this continuous distribution, which we call \textbf{rate distribution}, events are generated to populate and update the discrete distribution that we intend to sample. Knowledge about the rate distribution might come from theoretical knowledge about the underlying processes, direct observation, Monte Carlo simulations, etc. (For more details, see Section \ref{sec:assumptions}).

\paragraph*{An illustrative example}
To better understand the problem studied in this paper, let us first introduce an example from a real life application: a continuous-time simulation of Molecular-Beam-Epitaxial (MBE) growth on a metallic substrate \cite{Barkema1999,Breeman1996}, in the sub-monolayer regime. The set of energetically preferred positions of adatoms (atoms dropped on the surface) located on top of the metallic substrate forms a natural lattice with coordination number $z$ (i.e. $z$ denotes the number of neighbors of each site), typically a square lattice with $z=4$ or a honeycomb lattice with $z=3$. While new atoms are arriving on the substrate with a statistical rate determined by the beam intensity, the ones already present are hopping from one such preferred (lattice) position to a neighboring one, usually resulting in coalescence in islands. The hopping rate for an atom from site $A$ to a neighboring site $B$ depends on the atoms in the immediate vicinity of $A$ and $B$. If site $B$ is not occupied, the hopping rate is in very good approximation determined by the occupational state of the closest neighbors of $A$ and $B$. In the case of $z=4$, seen in Figure \ref{fig:atom}, this results in $2^{10}$ possible configurations, and, for each of these configurations, we can pre-compute the hopping rate \cite{Voter1986}. The simulation then proceeds by two steps:
\begin{itemize}
	\item[a)] the time is moved forward by a value $\Delta t$ equal to the inverse of the sum of all the rates of all possible events; 
	\item[b)]after this time increment, one event (hopping or arrival) is selected, with a probability proportional to its rate.
\end{itemize}
Therefore, we compute the rate of every possible move of every atom at every iteration and we sample a random event, employing a simple data structure: usually an array of size $zN$ which contains at every index the sum of the rates of all the events up to that one. A random number between zero and the sum of all the rates is generated and we move through the array until we reach a value larger than our random number and we sample that event. This might work well, but it does not scale as the sampling time grows linearly with the number of possible events. With limited literature search we find better structures for our problem, for instance, Complete Binary Trees (see Section \ref{sec:tree} and \cite{Wong1980} or, for a more flexible implementation, the Differential Search Tree from Maurer \cite{Maurer2017}), which sampling time grows logarithmically with the number of possible events, and even an optimal solution: the Alias method (see Section \ref{sec:alias}), proposed by Walker in 1974 \cite{Walker1974,Walker1977}, an ingenious method that, employing two tables of the same size as the number of possible events, allows constant time sampling, regardless their number or their rate. Alternatively, if we assume that all rates can be written as multiples of a unit, these can be stored in an array and sampled in constant time by picking a random site of the array; the obvious downside is the size of such array. While it can be compressed with the method from Marsaglia \cite{Marsaglia1963}, sampling from a compressed array requires $O \left( \log r_{max} \right)$, with $r_{max}$ the highest rate.

However, we see no significant improvement if we employ one of these structures. After each move, some of the configurations will have changed and we will have to rebuild the whole data structure from scratch, which costs a time that grows linearly with the number of possible moves $zN$, compromising the time saved with the sampling, even though only a limited number of possible moves have changed their rates. We can implement a (costly) update for the Complete Binary Tree (see Theorem \ref{thm:tree}) that requires $O(\log N)$ time, but that would still not scale well for larger numbers of atoms, and we would be tempted to optimize it in such a way that closer atoms are in the same branches, minimizing the number of operations required for the update, but as the atoms move they change neighbors, invalidating the optimization. 
%We also note that, since the time increment $\Delta t$ is the inverse of the sum of all rates, we can say that, if we assume the rates to be comparable, it decreases $\propto 1/N$. Even if we were able to sample events optimally, the computational time required for a fixed amount of simulated time still increases linearly with $N$.  

As the number of atoms necessary to study larger scale effects can be quite large, we would need a data structure that allows both optimal sampling and update of a random element. Unfortunately, and quite surprisingly, we were unable to find one for the general case. A quasi-optimal solution to the problem was given in 2003 by Matias et al.~\cite{Matias2003}. This method allows sampling in $O(\log^* N)$ time, with $\log^*$ the iterated logarithm, and the update of an arbitrary item in $O(2^{\log^* N})$ worst-case time and $O(\log^* N)$ amortized expected time. Unfortunately, the method of Matias et al.~is very complex to implement.A preliminary experimental study was done by van der Klundert \cite{Klundert2019}. Alternatively, the Acceptance-Rejection method (see section \ref{sec:AR}) does allow constant time updates, at the cost of performing samples in non-deterministic time, in which the expected value is dependent on the distribution of rates (see Theorem \ref{thm:ar}). Rajasekaran and Ross \cite{Rajasekaran} and Hagerup et al. \cite{Hagerup1993a} developed different solutions that allow for expected constant time updates and samplings by imposing restrictions on the updates that are not in general satisfied in our example or in similar settings, where the ratio between the largest and smallest rate can be quite large or even arbitrarily large.

Our example is not unique. Similar problems have been described not just in material physics, but also chemistry~\cite{Gillespie1977} and biochemistry~\cite{Slepoy2008}, and is in general relevant when we have an arbitrarily large number of possible events of known rate and their realization does not alter a significant fraction of them. It is therefore quite striking that we were not able to find in literature a general solution for such a relevant problem.

Since this is an intrinsically stochastic problem, it is sensible to ask whether the properties of the distribution of the rates of the possible outcomes, which can be determined either analytically or numerically assuming that the process that generates them is known, relates to the problem. An analysis from this point of view is also, to the best of our knowledge, missing in literature while there are some assumptions (see Section \ref{sec:assumptions}) that can be reasonable employed for large sets of applications that lead to some interesting solutions that we present in this paper.

\paragraph*{Our main contributions}
Our main contributions are twofold. First, we identify several cases where assumptions on the distribution and/or the number of events lead to expected constant time for sampling an event; while insertions and deletions of events can be done in amortized constant time. In particular, the known Acceptance-Rejection method gives expected constant time for non-increasing distributions; our new two-level Cascade method gives expected constant time for two large classes of distributions, and our new two-level Acceptance-Rejection method gives expected constant time regardless of the distributions. In several cases, the result only holds for a sufficiently large number of events; in all cases, bounds for the smallest and largest rate of events have to be known. Second, we give an experimental evaluation of several of the data structures, both from existing literature and those introduced in this paper.

\paragraph*{Organization of this paper.}
We start by defining our assumptions and the problem we are setting ourselves to solve (Section \ref{sec:assumptions}), then we will define and study the property of the data structures, both simple (Section \ref{sec:data-structures}) and multilevel (Section~\ref{section:multilevel}), that we employ to solve our problem. 
We perform an experimental analysis of our findings (Section \ref{sec:experiments}). Some conclusions are given
in Section~\ref{section:conclusions}.

\section{Problem Statement and Assumptions}
\label{sec:assumptions}

The data structures we study maintain an {\em Event Set} $\mathbb{E}$. The event set is a dynamic finite set (i.e. a finite set that can change over time). We call the elements of the Event Set {\em events}. Each event has a known, real, non-negative {\em rate}, that also can change over time; we denote the rate of event $e_i$ by $r(e_i)$. The rate of an event represents the number of expected occurrences in some arbitrary time unit.

Our data structures support as operations the insertion of an element (with a given rate), the deletion of an element, the change of the rate of an element, and a fourth operation: the {\em sampling } from the set of events. When we sample from the set of events, we randomly pick an event with a probability that is proportional to its rate. Thus, the probability that $e_i\in \mathbb{E}$ is sampled equals to
\begin{equation}
p(e_i) = \frac{r(e_i)}{\sum_{e\in \mathbb{E}} r(e) }.
\end{equation}
We make a further assumption, namely that we know the distribution of the rates of the events. More precisely, we have a probability density function $\rho$ such that the expected frequency of events with a rate between $a$ and $b$ equals $\int_a^b \rho(x) \, dx$. We assume that $\rho$ is known and fixed. This of course does not guarantee that at all times the possible events will be distributed following $\rho$, but that as the number of possible events $N \rightarrow \infty$ it will tend to $\rho$. It is useful to think of $\rho(r)$ as the continuous distribution from which the rates of the elements of the Event Set, the events that are possible at each given time, are sampled. We also assume that the rate has known and finite maximum $r_{max}$ and minimum $r_{min}$ (i.e., $\int_{r_{min}}^{r_{max}} \rho(x) \, dx=1$ and $\rho(r) = 0$ for $r \notin [r_{min},~ r_{max}]$). We finally also note that, by definition, rates of possible events are strictly positive, and therefore also $r_{min}$ and $r_{max}$ are defined as positive.

For the cases where $N$ is instead small, the Complete Binary Tree (see Section \ref{sec:tree}) is a good option, as it gives an $O(\log N)$ method that does not requires assumptions on the distributions of rates.

In many practical cases, the assumptions may be approximations of the real situation. Often, in such cases, the predicted expected times for our data structures can be good approximations of the true behaviour.

Given these assumptions, our problem is the following:

\begin{prob}
Given these assumptions, what is the most efficient method that allows for an event set $\mathbb{E}$:
\begin{itemize}
	\item sampling of an event (with each element selected with a probability that is proportional to its rate);
	\item update of the rate of an arbitrary number of events;
	\item removal or addition of an arbitrary number of events.
\end{itemize} 
\end{prob}

Our problem statement represents a not-so-uncommon type of problems in dynamic simulations where the processes are only locally interdependent, i.e. the realization of a process influences only up to a constant fraction of all possible processes. An update of an event can be implemented by deleting the event and inserting a new event with the new rate; in several cases, we thus do not discuss updates of rates separately.

\section{Data Structures}
\label{sec:data-structures}
In this section, we describe several data structures for the problem studied in this paper and briefly discuss dynamic arrays. After a short discussion of dynamic arrays, we review two basic well-known data structures: a Complete Binary Tree and the Acceptance-Rejection method. After that, we introduce three derivative methods which provide an efficient solution in different cases, depending on the probability distribution of the rates.

\paragraph*{Dynamic arrays}
In several cases, we store the events (or pointers to groups of events) as elements in an array. As we can add new elements to the data structure, the size of such an array can become too small. For this, we can use the standard data structure of {\em dynamic arrays}, also known as dynamic tables, see e.g.,~\cite[Chapter 17.4]{Cormen2001}. Several standard programming languages have this data structure built-in, e.g., dynamic arrays are provided under the name of \textit{vectors} in the C++ Standard Library. The main idea is that we use an array that is at least as large as needed, and copies all elements to an array of double size when the current array is too small. Occasionally, we have an operation that uses time, linear in the number of stored elements, but this happens infrequently, and the amortized time per insert (i.e., the total time divided by the number of operations) is bounded by a constant. For the details, we refer to e.g.~\cite[Chapter 17.4]{Cormen2001}.

\subsection{Complete Binary Trees}
\label{sec:tree}
A data structure that is commonly used for event sampling is the Complete Binary Tree. Here, a {\em Complete Binary Tree} is a binary tree (i.e., a rooted tree with each node having at most two children), with all levels completely filled, except possibly the lowest, which is partially filled. %is filled from the left. 
(Complete Binary Trees are also sometimes known as \textit{treaps}.) 
If we also impose that the lowest level is filled from the left, there is a simple implementation of Complete Binary Trees in arrays: we store the elements in an array $A[1 \cdots n]$, with the parent of node $A[i]$ being $A[ \lfloor {i/2} \rfloor ]$ ($i>1$). See e.g.~\cite[Chapter 6.1]{Cormen2001}, or \cite{EDELKAMP201289}.

\begin{figure}[!hbt]
\begin{center}
 \includegraphics[width=\textwidth]{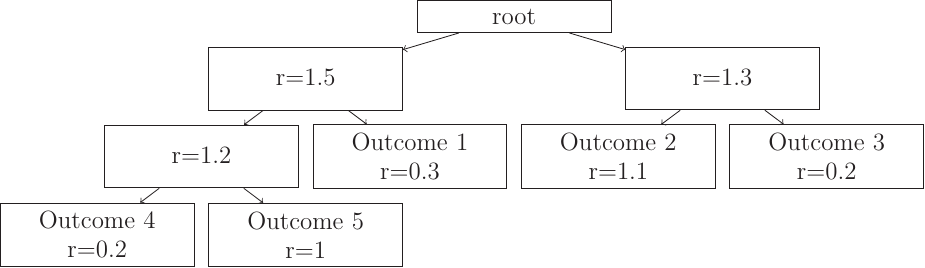}
 %\Description{Complete Binary Tree with five possible outcomes as leaves, each with a different rate. The three internal nodes' rates are equal to the sum of the rate of their children. The root of the tree is labelled as "root".}
 \caption{In the Complete Binary Tree, each node stores a variable called {\em rate}. Leaves, representing events, have the same rate as the corresponding event. Internal nodes have a {\em rate} equal to the sum of the rates of their children. An event is sampled by generating a random number between zero and the {\em rate} of the root (i.e. the sum of all rates): if this is smaller than the {\em rate} of the left node, we move to this node; otherwise, we subtract the left {\em rate} and move to the right node. This is repeated until we reach a leaf and the corresponding event is sampled.}
 \label{fig:tree}
 \end{center}
\end{figure}

While it would be tempting to group together in the same branch the events whose rate we might know to be correlated, for instance, the hopping rate of two spatially close atoms, we cannot assume that they will stay that way as the system evolves dynamically. The Complete Binary Tree has the advantage of an easier implementation, and it minimizes the average depth of the tree over all binary trees to $d = \lfloor \log_2 N \rfloor$, with $N$ the number of events.

 A schematic representation of the structure of a Complete Binary Tree is shown in Figure~\ref{fig:tree}. Each leaf represent an event and it is associated with its {\em rate}; internal nodes also have a {\em rate} associated with them and it is equal to the sum of the {\em rate}s of their children. Sampling is intuitive: a random number between zero and the sum of the rates of all the events ($r_{tot}$) is generated and, starting from the root, if this is smaller than the {\em rate} of the left node we move in that direction; otherwise, we subtract the {\em rate} of that node and we move to the right. This is repeated at most $d$ times until we reach a leaf. An update is performed by changing the {\em rate} of the corresponding leaf and updating the {\em rate} of the internal nodes between itself and the root. It is also possible to add or remove an event, by adding or removing a leaf with the usual methods, the {\em rate} of the affected internal nodes is updated. This is easiest in the array implementation: adding a new leaf just adds the element at the end of the array; in a deletion, we move the last element of the array to the position of the deleted element; in both cases, we update the rates of all nodes that are an ancestor of a replaced, inserted or deleted leaf. Under these assumptions, it is quite trivial to prove that all the operations that are interesting for us require logarithmic time. The following result 
can be easily derived from well-known insights and given here for completeness reasons.

\begin{thm}[Complete Binary Tree]
\label{thm:tree}
	Given an Event Set $\mathbb{E}$ of cardinality $N$ represented as a Complete Binary Tree:
	\begin{itemize}
		\item [(a)] the sampling of an event can be performed in $O(\log N )$ time;
		\item [(b)] the update of the rate of an event can be performed in $O(\log N)$ time;
		\item [(c)] the addition or removal of an event can be performed in $O(\log N)$ time.
	\end{itemize}
\end{thm}
\begin{proof}
	\begin{itemize}
		\item [(a)] The sampling of an event requires a number of operations
		 proportional to the number of nodes on the path between the root and the sampled leaf. For a Complete Binary Tree, this is at most $d = \lfloor \log_2 N\rfloor$~\cite{Black2019} and therefore it is $O(\log N)$.
		\item [(b)] In order to update the rate of an event, we perform a single operation on the leaf and then we update the internal nodes following backwards the same path as in (a). Therefore this is also $O(d) = O(\log N)$ operations.
		\item [(c)] First the leaf is deleted or added, which, for a binary heap, requires $O(\log N)$ time and then the rate of the nodes in the path from the deleted/added node to the root is updated. As we already mentioned, this costs also $O(\log N)$ time. 
	\end{itemize}
\end{proof}

\subsection{The Acceptance-Rejection method}
\label{sec:AR}
One of the classic methods is the Acceptance-Rejection method. Here, we have an array of size $N$ where each entry represents a possible outcome and its value is equal to its rate. Since the distribution is known, we assume that the maximum rate is also known.
\begin{figure}[hbt]
\begin{center}
 \includegraphics[width=.6\textwidth]{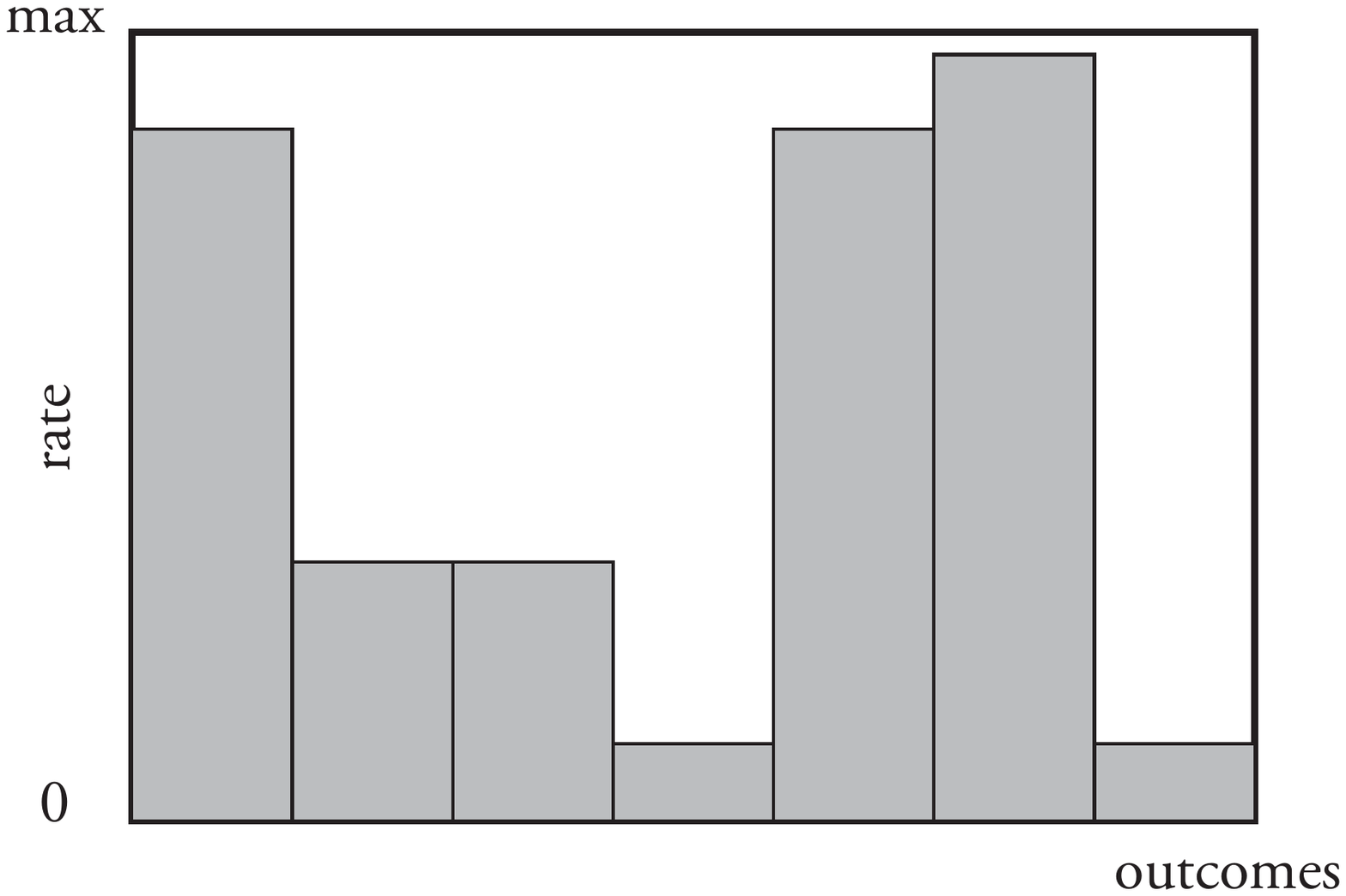}
 %\Description{An histogram where each possible outcome is represented as a bin with height representing its rate, comprised between 0 and max. Seven possible outcomes with different rates are shown.}
 \caption{Acceptance-Rejection structure. Each element of the array represents the {\em rate} of a possible event. An event is sampled by selecting a random element and generating a random number between zero and the $r_{max}$: if the latter is smaller than the former, the corresponding event is sampled; otherwise the process is repeated.}
\label{fig:ar-structure}
\end{center}
\end{figure}
 As no ordering is necessary, an element can be added and removed by simply adding or removing it from the array at any time, without any further preprocessing required. 
To sample an event, we randomly select an element and generate a random number between zero and the highest possible rate ($r_{max}$), which is known (see Section~\ref{sec:assumptions}); if this is larger than the value of the selected element, it is rejected and we draw a new one. Otherwise, it is accepted and sampled. The rate of an event is updated by simply changing the value of the corresponding element. A simple way to represent such data structure is as a histogram where each bin represents an element and their height is determined by their rate, up to the known maximum value $r_{max}$. An example of such representation can be seen in Figure~\ref{fig:ar-structure}.

We can make the data structure dynamic by using a dynamic array instead of a (usual) array; see the discussion at the start of this section. A new event can be added at the end of the array, and an element can be removed by moving the last element of the array to its position.
 
In contrast with other methods, the sampling time does not depend on the cardinality of the Event Set (i.e. the number of possible events $N$) while the updating time is always trivial, but we have to investigate how the rate distribution affects the sampling time. As this is a stochastic method, it is sensible to look at the expected time. We give a simple analysis of this method below.

\begin{thm}[Acceptance-rejection]
\label{thm:ar}
Given an Event Set $\mathbb{E}$ of cardinality N and largest rate $r_{max}$, represented as an Acceptance-Rejection structure:
\begin{itemize}
	\item[(a)] the sampling of an event can be performed in expected $O\left(\frac{r_{max}}{E[r]}\right)$ time, with $E[r]$ the expected value of the rate according to the distribution $\rho (r)$;
	\item[(b)] the update, addition or removal of an event can be performed in constant time. 
\end{itemize}
\end{thm}
\begin{proof}
\begin{itemize}
	%\item[(a)] The probability of sampling an event with outcome $r$ is given by the probability of selecting an event with such an outcome $\rho(r)$ multiplied by the probability of extracting a second number $s \leq r$:
	% Integrating over the whole :
  %\begin{equation}
	%P_{acc} = \frac{1}{r_{max}} \sum_{\mathbb{E}} \rho(r)\; r = \frac{E[r]}{r_{max}}\;.
	%\end{equation}
	 \item[(a)] The probability of selecting an event with rate $r$ is equal to the frequency of such events, which is expected to be $\rho(r)$. Then, a random number $S$ is generated from a uniform distribution with support $[0,r_{max}]$ and the event is accepted if $S \leq r$, the probability of which is $\frac{r}{r_{max}}$. We can then integrate it over all possible values of $r$ and write the probability of accepting an event of any rate as:
	\begin{equation}
	P_{sample} = \frac{1}{r_{max}}\,\int_{r_{min}}^{r_{max}} \, \rho(r) \, r \, dr = \frac{E[r]}{r_{max}}.
	\end{equation}
   
	Since this is a Bernoulli trial, the expected number of attempts before the first success is
	\begin{equation}
	E[n] = \frac{1}{P_{sample}} = \frac{r_{max}}{E[r]},
	\end{equation}
	and the number of operations is proportional to the number of attempts. 
	\item[(b)] The addition or removal of an event is performed by adding or removing an element to or from the vector. The rate of an event can be updated by simply changing the value of its element in the vector. All these actions require a constant number of operations, therefore they can be performed in constant time.
	\end{itemize}
\end{proof}
As the expected value of the rate cannot be smaller than the smallest possible rate, we can also say that
\begin{corollary}
\label{coroll:expAR}
	The sampling of an event can be performed in expected $O\left(r_{max}/r_{min}\right)$ time, with $r_{min}$ the smallest rate in the Event Set. 
\end{corollary}

\begin{figure}[!hbt]
\begin{center}
 \includegraphics[width=\textwidth]{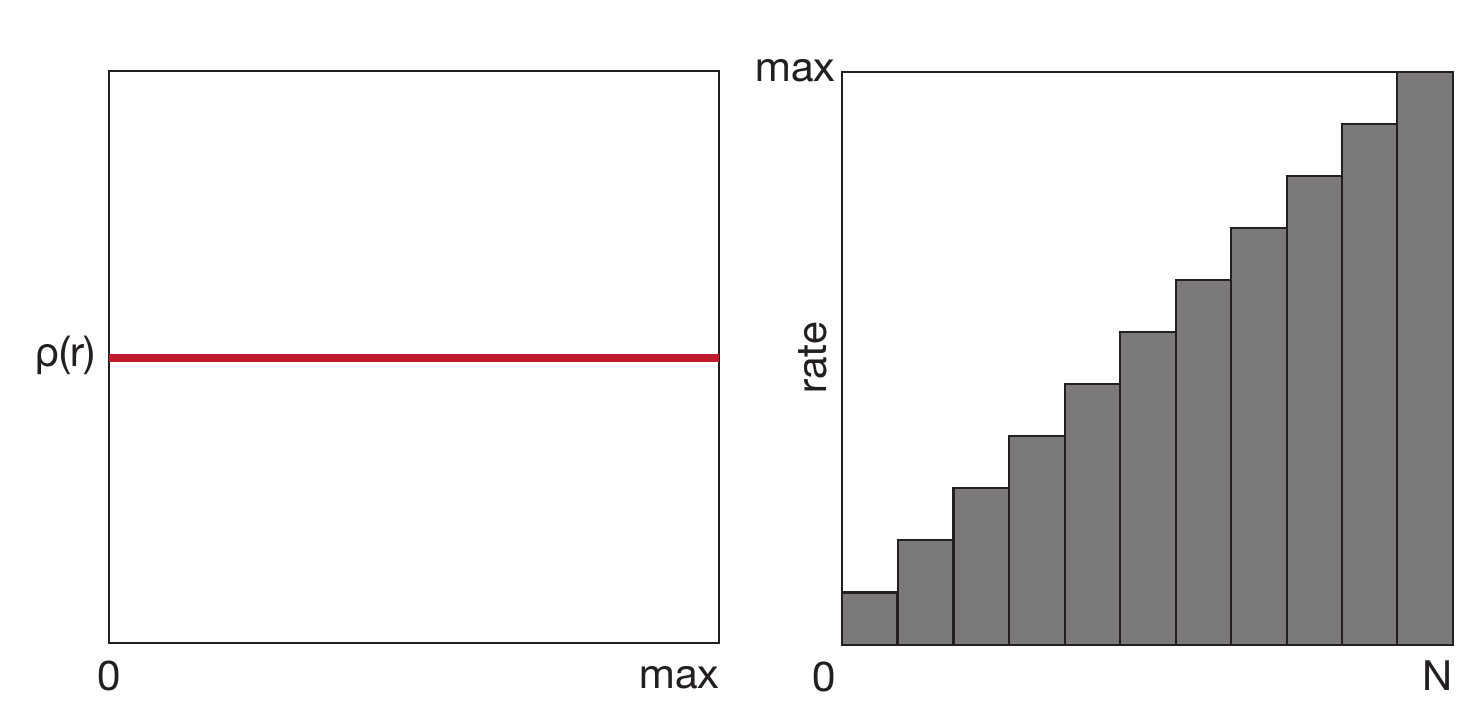}
 %\Description{On the left, a uniform rate distribution between zero and max. On the right, a histogram with eleven bins, each representing a possible outcome, with equally increasing height}
 \caption{On the left, a uniform rate distribution; on the right, the visualization of the corresponding Event Set. In order to avoid confusion in the following proofs, it is important to remember this distinction.
 Since order does not affect sampling, rates are ordered for clarity. }
\label{fig:ar-uni}
\end{center}
\end{figure}

To avoid confusion, we can visualize the Event Set $\mathbb{E}$ as a histogram of bins of equal width and height proportional to their rate, with the expected frequency given by the rate distribution $\rho(r)$. Note that, as we can see in Figure \ref{fig:ar-uni} the histogram does not look like the rate distribution. The sampling of an event is analogue to randomly shooting a dart on this area: if it lands inside a bin, that event is sampled; otherwise, it is rejected. 

Let us step back to the result of Theorem \ref{thm:ar} for sampling. We can easily imagine a worst-case, where all the events except one have a rate arbitrarily smaller than the largest and the sampling time, therefore, grows arbitrarily, and a best-case, where all the events have the same rate and the sampling time is constant. Is there a more general assumption we can introduce on the rate distribution $\rho(r)$ that would still guarantee expected constant time? We will show that assuming that the rate distribution is non-decreasing is sufficient to guarantee expected constant time.

The probability of selecting an outcome with a given rate is proportional to the number of elements with that rate. We can therefore write
\begin{equation}
\label{eq:defMedian}
	\int_{r_{min}}^{\bar{r}} \, \rho(r) \; dr= \int_{\bar{r}}^{r_{max}} \, \rho(r) \; dr,
\end{equation}
with $\rho(r)$ the rate distribution and $\bar{r}$ the median of the distribution $\rho(r)$~\cite{loeve1977median}, i.e. the real number for which 
\begin{equation}
\int_{r_{min}}^{\bar{r}} \, \rho(r) \; dr = \frac{1}{2} \hspace{10mm} \int_{\bar{r}}^{r_{max}} \, \rho(r) \; dr = \frac{1}{2},
\end{equation}
which is guaranteed to be unique if $\rho(r) > 0$ in the open interval $(r_{min}, r_{max})$. Since the possible outcomes are selected from an uniform distribution, this implies that the probability of selecting a possible outcome whose rate is at least $\bar{r}$ or larger is
\begin{equation}
\label{eq:p_select_half}
  P_{select} (r \geq \bar{r}) \geq \frac{1}{2}.
\end{equation}

\begin{figure}[!hbt]
\begin{center}
 \includegraphics[width=.8\textwidth]{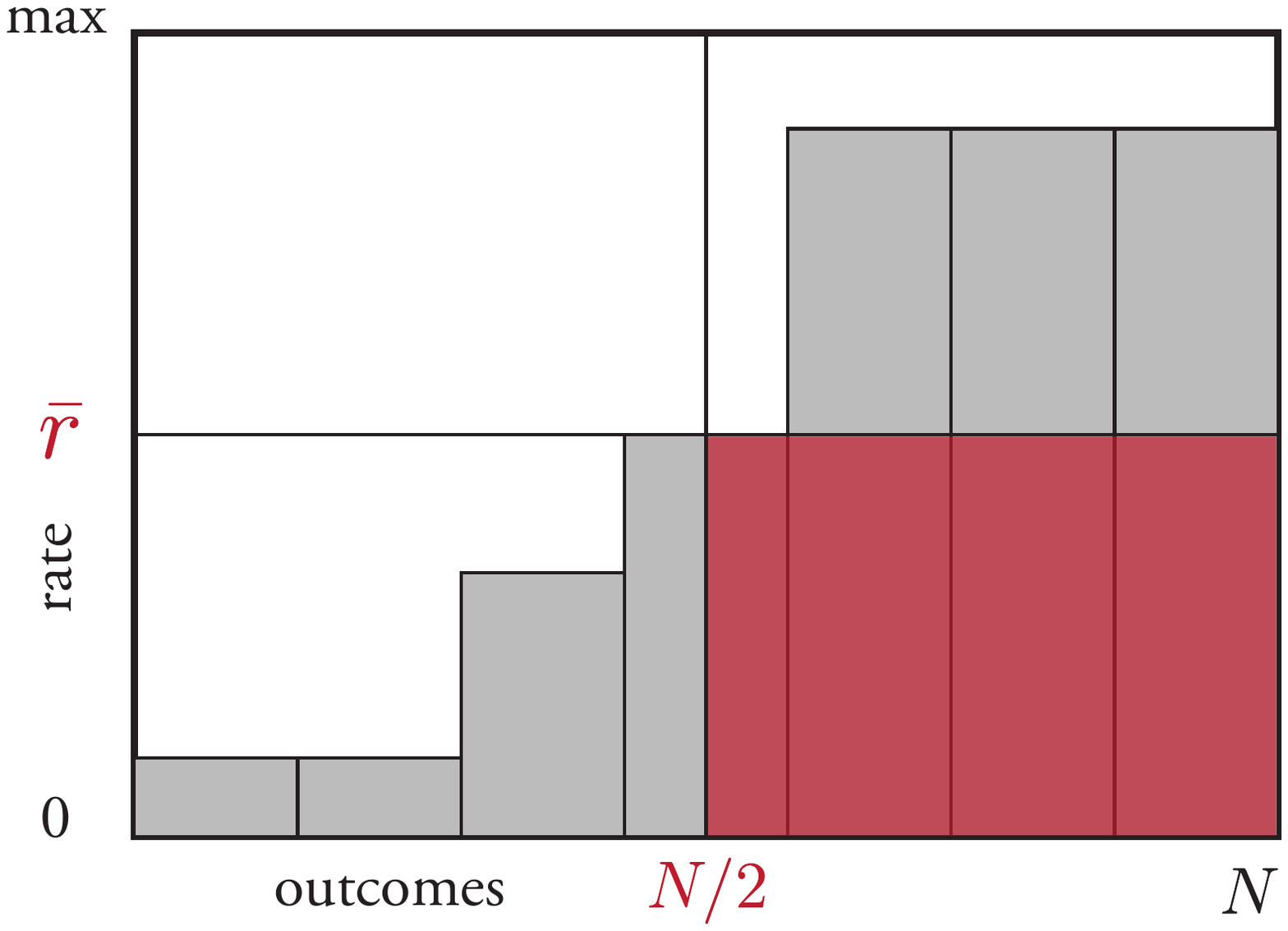}
 %\Description{A histogram with seven bins representing possible outcomes from a non-decreasing rate distribution. On the two axis, the midpoint of the outcomes ($N/2$) and the median ($\bar{r}$) are clearly labeled. The area with rate comprised between 0 and $\bar{r}$ and outcomes between $N/2$ and $N$ is also highlighted.}
 \caption{Each possible outcome is represented as a rectangle with unit width and height proportional to its rate. Since order does not affect sampling, rates are ordered for clarity. The probability of sampling an event is equal to the ratio between filled and total space in such a representation. From this geometric argument we can prove that, for a non-decreasing rate distribution, the probability of sampling an outcome is at least $\frac{1}{4}$ due to Lemma \ref{lemma:rbar}.}
 \label{fig:graphAR}
 \end{center}
\end{figure}

\begin{lemma}
\label{lemma:rbar}
	If the rate distribution $\rho(r)$ is a non-decreasing function of $r$, its median is at least the middle of the interval $[r_{min},~ r_{max}]$ (i.e., at least $\displaystyle \frac{r_{min}+r_{max}}{2}$). 
\end{lemma}
\begin{proof}
	First, we rewrite Equation \ref{eq:defMedian} as
	\begin{equation}
			\int_{r_{min}}^{\bar{r}} \, \rho(r) \; dr- \int_{\bar{r}}^{r_{max}} \, \rho(r) \; dr = 0.
	\end{equation}
	Using that $\rho(\bar{r}) \geq \rho(r)$ for each $r\in [r_{min},\bar{r}]$ and
	$\rho(\bar{r}) \leq \rho(r)$ for each $r\in [\bar{r},r_{max}]$, as we assume that $\rho$ is a non-decreasing function, it follows that
	\begin{equation}
		\rho(\bar{r}) \int_{r_{min}}^{\bar{r}} dr - \rho(\bar{r}) \int_{\bar{r}}^{r_{max}} dr \geq 0,
	\end{equation}
	assuming that $\rho(\bar{r})$ is non-zero. Finally,
	\begin{equation}
		\label{eq:bar-r}
		\bar{r} - r_{min} - r_{max} + \bar{r} \geq 0
	%\end{equation}
	%can also be written as
	%\begin{equation}
        \;\;\Rightarrow\;\;
		\bar{r}\geq \frac{r_{max}+r_{min}}{2}.
	\end{equation}
\end{proof}
We are ready to prove the following theorem:
\begin{thm}
\label{th:ARNonDecr}
	An Acceptance-Rejection structure with a non-decreasing rate distribution performs sampling of a possible outcome in expected constant time.
\end{thm}
\begin{proof}
  Since the rates of possible events are strictly positive, we can write Equation~\ref{eq:bar-r} as:
	\begin{equation}
		\bar{r}\geq \frac{r_{max}+r_{min}}{2} \geq \frac{r_{max}}{2}.
	\end{equation}
	Remembering from the proof of Theorem~\ref{thm:ar}a, we can write the probability of accepting an outcome with rate $r$, assuming that an outcome with rate $r \geq \bar{r}$ is already selected, is
	\begin{equation}
		P_{accept}(r\, |\, r \geq\bar{r}) = \frac{r}{r_{max}} \geq \frac{\frac{r_{max}}{2}}{r_{max}} = \frac{1}{2}.
	\end{equation}
	Remembering the result of Equation~(\ref{eq:p_select_half}), the probability of successfully sampling an outcome with $r\geq \bar{r}$ therefore is
	\begin{equation}
		P_{sample}(r\geq\bar{r}) = P_{select}(r\geq\bar{r}) \; \cdot P_{accept}(r \,|\, r\geq\bar{r}) \geq \frac{1}{2} \cdot \frac{1}{2} = \frac{1}{4}.
	\end{equation}
The probability of successfully sampling an outcome from a subset of the Event Set cannot be larger than the probability of sampling an outcome from the entire Event Set, which puts an upper boundary on the expected number of attempts before sampling an outcome 
	\begin{equation}
		E[t_{sample}] = \frac{1}{P_{sample}} \leq \frac{1}{P_{sample}(r\geq\bar{r})} = 4 = O(1),
	\end{equation}
	with $P_{sample}$ the probability of successfully sampling an outcome from the entire Event Set. 
\end{proof}
A visualization of this proof can be seen in Figure \ref{fig:graphAR}. This is a very powerful result: such a simple method allows
constant time sampling for any Event Set with a non-decreasing rate distribution.
 \subsection{The Alias Method}
 \label{sec:alias}
The Alias method, introduced by Walker \cite{Walker1974,Walker1977} is a very ingenious solution to the static case of our problem. Each event is conceptually stored in a "bucket" of size $r_{tot}/N$; if a bucket is not already full, the remaining space is assigned to another event, denoted as its \textit{alias}, that is overfilling its bucket. The rate that has been assigned to the alias is then removed from its original bucket. This is repeated until each bucket is exactly full. 

The buckets are represented as an array of size $N$, each element storing the fraction of the bucket assigned to the alias. To sample an event, an element and a random number between zero and one are generated. If this is larger than the value stored in the element, the corresponding event is sampled; otherwise, we sample its alias.   

As the number of steps required for sampling is fixed, the time required is constant. Unfortunately, except for some very particular cases, any update would be extremely costly and it would often require a complete rebuild of both tables, which takes at least $O(N)$ time. Nevertheless, we are presenting this method both for completeness and as a potential building block for multilevel methods. 

\section{Multilevel Methods}
\label{section:multilevel}
As we have seen, the Acceptance-Rejection method works better when the possible outcomes have a limited range of rates; if this is not the case, we can split the Event Set in multiple groups according to their rate, use one of the other methods to sample a group, and then the Acceptance-Rejection method to sample an element from that group~\cite{Slepoy2008}. 
We call such combinations of different methods \emph{multilevel methods} and the structure that stores the groups \emph{superstructure}. In this section, we present some of these combinations that have very powerful proprieties that will be shown in the next section.

\subsection{Exponential grouping}
\label{subsection:grouping}

All our two level methods employ the same data structure for the lower level.

The events are grouped according to their rates. Each group consists of all events with a rate in a specific interval. The sizes of these intervals grow exponentially, and hence we will refer in the successive subsection this grouping method by {\em exponential grouping}.

Fix some constant $c>1$. A typical example would be to take $c=2$. Different choices for $c$ can affect the 
constant factors of the running time: larger values of $c$ would slow down selection in the lower level of the data structure, but could speed up selection in the upper level of the data structure.

Number the groups starting at $1$. The group with index $i$ consists of all events with rate $r$ in
the interval 
\begin{equation}
r \in \left[ c^{i-1} \cdot r_{\min}, \min \left\{ c^i \cdot r_{min}, r_{max} \right\} \right),
\end{equation}
adding the value $r_{max}$ to the last group, (i.e., all intervals except the last are right-open.)

For each group, we use a separate Acceptance-Rejection data structure to sample an event.

\begin{lemma}
After a group is selected, sampling an event from that group can be done in $O(c)$ expected time.
\label{lemma:sampleingroup}
\end{lemma}

\begin{proof}
Note that the ratio between the largest and smallest rate of events from one group is bounded by
$\frac{c^{i} \cdot r_{\min}}{c^{i-1} \cdot r_{\min}} = c$. Thus, the expected number of `rounds' of the
Acceptance-Rejection method until an event is selected from the group is bounded by $c$, which we assumed to be
a constant.
\end{proof}

Updating rates, inserting new events, and deleting events in the lower level data structure
all can be done in constant time. An update can be
performed by deleting the event with the old rate, and inserting an event with the new rate. We fix an array
with an element for each group, that points to the Acceptance-Rejection data structure of that group. If we
insert an element, with a constant number of arithmetic operations, we can determine its group, find the corresponding Acceptance-Rejection data structure, and add the event. To delete an event, we need a pointer to its location in
its Acceptance-Rejection data structure, and delete it as in Theorem~\ref{thm:ar}.

What remains is to build data structures to sample a group, where we need to select each group with
a probability that is proportional to the total rate of all events in the group. For this, we have for 
each group a variable that maintains this total rate of all events in the group. Apart from that,
we have different method to sample groups, which are discussed in the successive subsections.

\subsection{Tree of Groups}\label{sub:treeOfGroups}
Let us assume that the Event Set has an arbitrarily large cardinality but the range of rates is such that the number of groups required to cover it is limited. In such a case we can employ a Complete Binary Tree as a superstructure and obtain a very useful result: both update and sample are performed in $O\left(\log \log \frac{r_{max}}{r_{min}}\right)$ expected time. While this is not constant time, it is very small without requiring any further assumption on the rate distribution. A similar method, called SSA-CR (Stochastic Simulation Algorithm - Composition and Rejection), was introduced in~\cite{Slepoy2008}.

The method thus works as follows. We group the events by the exponential grouping method (see Section~\ref{subsection:grouping}).
% : Events are grouped according to their rate; the group with index $i$ includes all the events with rate
% \begin{math}
% 	\frac{r_{max}}{k^{i-1}} < r \leq \frac{r_{max}}{k^i}
% \end{math}
% with $k$ a positive constant. 
Each group is represented both as a leaf of a Complete Binary Tree (see Section \ref{sec:tree}), whose rate is given by the sum of the rates of all the events in the group, and as an Acceptance-Rejection structure where all its events are stored.  
This total rate can easily be maintained under insertions, deletions and updates; after such an operation the difference is added or subtracted from the group rate.

To sample an event, we first sample a group from the Complete Binary Tree in the previously described way (see Section \ref{sec:tree}) and the Acceptance-Rejection sampling (see Section \ref{sec:AR}) is performed inside it. Updates are trivial unless they require events to be moved to a different group; in that case, the relative element is removed from its group and added to the new one. 

\begin{thm}[Tree of Groups]
\label{th:tog}
	Given an Event Set $\mathbb{E}$ represented as a Tree of Groups:
	\begin{itemize}
		\item [(a)] the sampling of an event can be performed in $O\left(\log \log \frac{r_{max}}{r_{min}}\right)$ time;
		\item [(b)] the update, addition or removal of an event can be performed in $O\left(\log \log \frac{r_{max}}{r_{min}}\right)$ time.
	\end{itemize}
\end{thm}
\begin{proof}
\begin{itemize}
	\item[(a)] Since the groups are stored in a Complete Binary Tree, the time to select a group grows logarithmically with the number of groups; the lower boundary of the i-th group is, by definition, $\frac{r_{max}}{k^i}$; and the general lower boundary is $r_{min}$, we can write the number of groups $n$ as
\begin{equation}
	\frac{r_{max}}{k^n} = r_{min} \; \; \Rightarrow \; \; n = \log_k\left(\frac{r_{max}}{r_{min}}\right).
\end{equation}
Therefore, the time required to select a group from the Complete Binary Tree is $O\left(\log n\right)=O\left(\log \log \frac{r_{max}}{r_{min}}\right)$. 
Once we have selected a group, sampling an event from the group uses expected constant time (Lemma~\ref{lemma:sampleingroup}.)
% its boundaries are known. The expected number of attempts before successfully sampling an outcome from a group with index $i$ is, according to Corollary \ref{coroll:expAR},
% \begin{equation}
% 	E[t_{sample, i}] \leq \frac{\frac{r_{max}}{k^{i-1}}}{\frac{r_{max}}{k^i}} = k = O(1),
% \end{equation} 
% since $k$ is a constant. 
\item[(b)] Adding, removing and updating an event inside its group is performed in constant time (see Theorem \ref{thm:ar} and Section~\ref{subsection:grouping}). In order to maintain consistency it is necessary to update the rates in the Complete Binary Tree, which is performed in the same time as a sampling (see Theorem \ref{thm:tree}). 
\end{itemize}
\end{proof}

\subsection{Cascade of Groups} 
\label{sec:cascade}
We have previously shown that the Acceptance-Rejection is optimal for any non-decreasing rate distribution. While we would like to find a similar result for all decreasing rate distribution, therefore completing the solution for the general problem, we will split them into different subsets and attack them one at the time. Let us first consider those rate distributions that, according to some definition, decrease fast enough. For such rate distributions, most of the events will have lower values of rate; we must therefore store groups in a way that prioritize events with a lower rate. 

% Events are grouped according to their rate; the group with index $i$ includes all the events whose rate is in the interval $[ c^i \cdot r_{min}, c^{i+1} \cdot r_{min} ) $, with $c$ a (real) constant larger than 1. 
Again, we use exponential grouping, see~\ref{subsection:grouping}.

In the analysis below, we assume that $r_{max}$ is a multiple of $r_{min}$. If this is not the case,
we can have a slightly smaller last group. It is easy to see that the difference in expected
running time is bounded by a constant. 
%We also ignore the event with probability exactly $r_{max}$ --- if present, it can be added to the last group.

\begin{figure}
  \centering
  \includegraphics{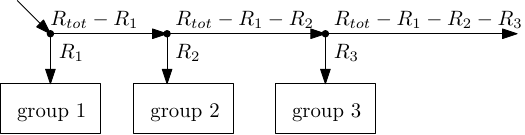}
  \caption{The first three groups in the Cascade of Groups structure. The numbers give the expected proportion
  of times the arrow is followed.}
  \label{figure:cascade}
\end{figure}

Our data structure is as follows.
We have a linked list ~\cite[Chapter 10.2]{Cormen2001} with an element for each group, which has both a pointer to its Acceptance-Rejection data structure and the value of the sum of all the rate in the group (denoted, for the $i$th group, as $R_i$). See Figure~\ref{figure:cascade} for visualization of such superstructure. 

%TO DO: CHANGE r_1 r_2 .... to R_1 R_2 .... DONE

%Each group is represented both as an element in a linked list~\cite[Chapter 10.2]{Cormen2001} and as an Acceptance-Rejection structure where all its events are stored. 

Sampling is, again, in two phases. A random number $rand\in[0,R_{tot}]$ is generated, with $R_{tot} = \sum R_i$; if $rand$ is larger than the sum of the rates in the first group $R_1$, this is subtracted from $rand$ and we move to next group. This is repeated until a group is selected. A sample from the selected Acceptance-Rejection structure is then performed. Updates, addition or removal of events are performed inside the groups following the methods described in Section \ref{sec:AR}; $R_{tot}$ and the sum of the rates in the involved group (or groups, for an event that changes group after an update) are also updated. 

Let us start by introducing this useful Lemma for the Cascade of Groups:
\begin{lemma}
In a Cascade of Group, if there is a constant $\alpha <1$ such that, for each group, the expected sum of rates of a group is at most $\alpha$ times the expected rate of the previous group, then the expected time to select a group is $\displaystyle O\left(\frac{1}{1-\alpha} \right)= O(1)$.
\label{lemma:cascade}
\end{lemma}

\begin{proof}
Once we reach the $i$th group in the Cascade of Group, the expected probability of selecting it is the expected rate of that group divided by the sum of the expected rate of that and all of the following groups. We call the expected rate of the $i$th group $E[R_i]$. 
Supposing we have $g$ groups, the expected total rate of the $i$th group and all following groups is at most
\[
 E\left[\sum_{j=i}^{g}R_j\right] = \sum_{j=i}^{g}E[R_j] \leq \sum_{j=i}^{g} E[R_j] \cdot \alpha^{j-i} \leq 
 \sum_{j = i}^{\infty} E[R_j] \cdot \alpha^{j-i}
 \]
 \begin{equation}
 = E[R_i] \cdot \sum_{j=0}^{\infty} \alpha^{j} =
\frac{E[R_i]}{1-\alpha}.
\end{equation}
The expected probability that when we are at a group $i$, take an element from that group is thus here at least
\begin{equation}
\frac{E[R_i]}{E[R_i] \cdot \left({\frac{1}{1-\alpha}}\right)} = {1-\alpha}.
\end{equation}
We can view the execution of the algorithm as an experiment that is repeated till the first success; with each round, we have a probability of success that is at least $1-\alpha$. The expected number of steps before selecting a group is thus at most $\displaystyle O \left(\frac{1}{1-\alpha}\right) = O(1)$.
\end{proof}

Following this Lemma, we can prove that the Cascade of Groups is a constant time solution if the rate distribution decreases fast enough:

\begin{thm}
Suppose we have a constant $c>1$ such that for all $r \in [r_{min}, r_{max}/c]$: 
\begin{equation}
 \rho(cr) \leq \rho(r)/c^\beta,
 \end{equation}
 with $\beta > 2$, then the Cascade of Groups data structure gives expected constant time to sample an event.
 \label{theorem:cascade}
\end{thm}
\begin{proof}

We first relate the expected total rate of group $i$, $E[R_i]$, with the expected total rate
of group $i-1$, $E[R_{i-1}]$:
\begin{equation}
\begin{split}
  E[R_i] & = \int_{c^i \, r_{min}}^{c^{i+1} \, r_{min}} r\;\rho(r) \; dr =
     c \int_{c^{i-1} \, r_{min}}^{c^{i} \, r_{min}} cr'\;\rho(cr') \; dr' \\
  & \leq 
  c \int_{c^{i-1} \,r_{min}}^{c^i\, r_{min}} 
  \frac{c}{c^\beta} r'\;\rho(r') \; dr' = c^{2-\beta} E[R_{i-1}],
\end{split}
\end{equation}
which follows by using the 
substitution
$r' \rightarrow r/c$. Thus, we get the result by Lemma \ref{lemma:cascade}, and noting that $c^{2-\beta}<1$, when $\beta>2$.

Once inside a group, sampling takes expected constant time by Lemma~\ref{lemma:sampleingroup}.
% \begin{equation}
%   E[t_{accept}] = O \left(\frac{c^{i+1} \; r_{min}}{c^i \; r_{min}}\right) = O(c) = O(1).
% \end{equation}
\end{proof}

\subsection{Reversed Cascade of Groups}
\label{sec:rev-cascade}
We can obtain a similar result when rates decrease sufficiently slow, just by reversing the superstructure. In the Reversed Cascade of Groups, we use exponential grouping (see Section~\ref{subsection:grouping}), and again place these in a linked list (as for the Cascade of Groups), except that we link the groups in the reversed order, i.e., we start with the group with the events with the largest rate. Thus, if we have $g$ groups, we first decide if we sample an element from $g$th group, then from $g-1$th group, etc.

\begin{thm}
Suppose we have a constant $c>1$ such that for all $r \in [r_{min}, r_{max}/c]$: 
\begin{equation}
 \rho(cr) \geq \rho(r)/c^\beta,
 \end{equation}
 with $\beta < 2$, then the Reversed Cascade of Groups data structure gives expected constant time to sample an event.
 \label{theorem:reversecascade}
\end{thm}
\begin{proof}
Let $i<g$, i.e., group $i$ is not the group with the events with largest rates.
We again relate the expected total rate of group $i$, with the expected total rate
of group $i-1$. By substitution $r' \rightarrow r/c$, we obtain:
\begin{equation}
\begin{split}
  E[R_i] & = \int_{c^i \, r_{min}}^{c^{i+1} \, r_{min}} r\;\rho(r) \; dr =
     c \int_{c^{i-1} \, r_{min}}^{c^{i} \, r_{min}} cr'\;\rho(cr') \; dr' \\
  & \geq 
  c \int_{c^{i-1} \,r_{min}}^{c^i\, r_{min}} 
  \frac{c}{c^\beta} r'\;\rho(r') \; dr' = c^{2-\beta} E[R_{i-1}].
\end{split}
\end{equation}
All groups, except group $g$, thus fulfill the condition of Lemma~\ref{lemma:cascade}. Visiting
the first group costs constant time, and thus, with Lemma~\ref{lemma:cascade}, and because
$c^{\beta-2}<1$ here,
we see that the
expected time to select a group is bounded by a constant. Again, the sampling inside a group costs
expected constant time (Lemma~\ref{lemma:sampleingroup}).
\end{proof}

%We finally note that selecting the value of $c$ has practical consequences, $E[t_{sel}]$ decreases while $E[t_{accept}]$ increases with $c$, but for most uses $c=2$ can be a reasonable compromise. 

We now have optimal solutions for small numbers of events (Tree of Groups, see Section \ref{sec:tree}), small range of rates (Tree of Groups, see Section \ref{sub:treeOfGroups}), non-decreasing rate distributions (Acceptance-Rejection, see Section \ref{sec:AR}), fast decreasing rate distributions (Cascade of Groups, see Section \ref{sec:cascade}) and slow decreasing rate distributions (Reverse Cascade of Groups, see Section \ref{sec:rev-cascade}). In the next Section, we will introduce an optimal solution of our problem for any rate distribution, if the number of events is significantly large.

\subsection{Two Levels Acceptance-Rejection}
\label{sub:2lar}
We now discuss a Two Levels structure where both levels use the Acceptance-Rejection method.
%We give two versions: a heuristic version that performs well in practice and a theoretical result that gives expected constant time regardless of the distribution.

Again, we group the events with exponential grouping (see Section~\ref{subsection:grouping}). 

The elements of the top-level Acceptance-Rejection structure are called {\em bins}. Each bin has a rate, and points to a group.
We allow that groups have multiple bins, and the total rate of all the bins of a group equals the total rate
of all events of the group.

By using multiple bins per group, we can obtain a constant expected time to sample an event,
regardless of the rate distribution, and amortized constant time for insertions and deletions.
However, insertions and deletions can require multiple pointer operations and can be slow in practice.

Suppose we are given the values of $r_{min}$ and $r_{max}$. We now choose a value $B \geq r_{max}$.
and $c>1$ that we consider to be constants. 
%
% Again, we partition the range $[r_{min},r_{max}]$ in groups $[c^{i-1} \cdot r_{min}, \min \{c^i \cdot r_{min},r_{max}\})$, and add $r_{max}$ to the last one. 
Let $g = \log_c \left\lceil \frac{r_{max}}{r_{min}}\right\rceil$ be the number of groups we obtain by using
exponential grouping.

%For each group, we have an Acceptance-Rejection data structure, which is used to sample an event from a group, once the group is selected. As the ratio between the largest and smallest rate in a group is at most $c$, we can sample an element from a group in $O(c)$ expected time.

Now (as described in Section~\ref{subsection:grouping}), each group at the lower level uses an Acceptance-Rejection data structure to sample an event from the group, but also the superstructure is
 an Acceptance-Rejection structure, where the bins play the role of events.  Each group has at least one bin in the superstructure. Each bin has associated with it a non-negative real number, called {\em value}. For each group, all its bins have value $B$, except possibly the last (or the only) bin of the group.

A group is selected by randomly drawing a bin and a real number between $0$ and $B$. If this random number is at most the value of the bin, then we select the bin and the group associated with it. Otherwise, we repeat this operation until a bin is selected. It is easy to see that the probability to select a group is proportional to the sum of the values of its bins, which is equal to the rate of the group.

The bins of a group have a pointer to the previous and next bin of the same group, and to an object that represents the group. That object has a pointer to the last bin of the group, a local variable equal to the rate of the group, and a pointer to the Acceptance-Rejection structure of the group.

An element can be added by inserting it in the Acceptance-Rejection structure of its group and then adding its rate to the value of the last bin of the same group. If this becomes larger than $B$, say it becomes $x > B$, then we create a new bin for the group, add it to the superstructure, set the value of the now second to last bin of the group to $B$, set the value of the new last element of the group to $x-B$, and set the pointers to and from the last and second to last bin of the group correctly. 

Deleting an element is done by deleting it from the Acceptance-Rejection structure of its group, and subtracting its rate from the last bin of the group. Suppose the rate of this last bin becomes $y$. If $y$ is positive, it is simply updated. Otherwise, we delete the last bin of the same group and decrease the rate of the bin that has become the last of the group to $B+y$.

Note that these operations ensure that all bins have a non-negative value that is at most $B$ and that insertions and deletions involve a constant number of operations, and thus cost amortized constant time.

\begin{lemma}
Suppose the total rate of all events is $R_{tot}$, and we have $g$ groups. Then, the expected time to
select a group is $O(1 + \frac{Bg}{R_{tot}})$.
\end{lemma}

\begin{proof}
We have $m \geq g$ bins. All bins have a value at most $B$, and at most $g$ bins have a value smaller than $B$, so we have that $R_{tot} > B(m-g)$.
The expected value of a bin in this structure can be written as $E[b] = \frac{1}{m}\sum^1_{i=m} b_i = {R_{tot}}/{m}$,
with $b_i$ the value of the $i$-th bin. 
We now can apply Theorem~\ref{thm:ar}. As the event in this step is the selection of a bin, the expected value of a bin ($E[b]$) plays the role of $E[r]$, and the maximum value of a bin ($B$) plays the role of $r_{max}$, so, by Theorem~\ref{thm:ar}, the expected
time to select a bin and therefore a group is $O(\frac{B}{R_{tot}/m}) = O(Bm / R_{tot})$.
Now, observe that
\begin{equation}
    \frac{Bm}{R_{tot}} = \frac{Bm-Bg}{R_{tot}} + \frac{Bg}{R_{tot}} < \frac{R_{tot}}{R_{tot}}+ \frac{Bg}{R_{tot}} = 1+ \frac{Bg}{R_{tot}}.
\end{equation}
The lemma now follows.
\end{proof}

Thus, recalling Lemma~\ref{lemma:sampleingroup} and that $g = \lceil \log_c \frac{r_{max}}{r_{min}} \rceil$, we can state the following.
\begin{thm}
\label{th:2AR-MB}
The Two Levels Acceptance-Rejection method, with $B\geq r_{max}$, $c>1$, allows to perform
\begin{itemize}
  \item insertions and deletions in amortized constant time, and
  \item sampling in expected time 
  \begin{equation}
  O\left( c + \frac{B \cdot \log_c \frac{r_{max}}{r_{min}}}{R_{tot}} \right),
  \end{equation}
\end{itemize}
when $R_{tot}$ is the total rate of all current events in the data structure. 
\end{thm}
Interestingly, this means that the Two Levels  Acceptance-Rejection method allows  for constant time sampling if $R_{tot} > B \cdot g$. Therefore if there are enough events in the structure to satisfy this condition, we have a method than can be applied to any rate distribution, as long as $r_{max}$ and $r_{min}$ are known.

Two parameters can be set that influence the expected time, namely $B$ and $c$. When we increase $B$, we can expect fewer operations that create or delete a bin, and thus would decrease the time needed for
pointer operations, but it also means that the term $\frac{Bg}{R_{tot}}$ is larger, thus making the data structure viable only for larger total rate values $R$. When we increase $c$, we have fewer groups (as
$g$ is $\left\lceil \log_c \frac{r_{max}}{r_{min}}\right\rceil$), which decreases the time in the superstructure, but it increases the time to sample inside a group. 

\section{Experimental Analysis}
\label{sec:experiments}
We have performed an experimental analysis of the sample time of the previously described methods in order to confirm our asymptotic findings. We implemented these methods in the C++ language, building on top of C++11 Standard Library (in particular \textit{random}). We computed the expected sample and the update time of each data structure by recording the time required with the \textit{high\_resolution\_clock} method of the \textit{chrono} library. Average and variance are computed according to the Walford method~\cite{Welford1962}. Our implementation is available in a GitHub repository\footnote{\url{github.com/federicodambrosio/dynamic-sampling-code}}.

Since rates can be re-scaled to different time units, the maximum rate is fixed to $1$ in some arbitrary units, while $r_{min}$, the minimum possible rate, is a controllable parameter, together with $N$, the number of possible events, and the rate distribution. 

We have included 5 monotonic rate distributions:
\begin{enumerate}
    \item an increasing distribution, $\rho(x) = k * x$;
    \item a uniform distribution, $\rho(x) = k$;
    \item a decreasing distribution with $\beta < 2$, $\rho(x) = k/x$;
    \item a decreasing distribution with $\beta = 2$, $\rho(x) = k/x^2$;
    \item a decreasing distribution with $\beta > 2$, $\rho(x) = k/x^3$;
\end{enumerate}
with $k$ the appropriate normalization constant. We set the constant for exponential grouping (see Section~\ref{subsection:grouping}) to $c = 2$ in all the Multilevel Methods. For each distribution, we vary the values of $r_{min}$ and $N$ and generate $100$ random Event Sets for each of them. On each Event Set, we perform $10^4$ samplings, $10^4$ updates only on the Two Levels Acceptance-Rejection and $100$ updates on all the others, and we compute the average of the CPU time required over all of them. The code is compiled and executed on the following system:

\begin{tcolorbox}
Processor: AMD Ryzen 5 3600X\\
RAM: 16 GB\\
Storage:  Crucial P2 1 TB M.2-2280 NVME Solid State Drive\\
Compiler: gcc 9.3.0.
\end{tcolorbox}

\subsection{Complete Binary Trees}
\begin{figure}[htbp]
\begin{center}
 \includegraphics[width=.7\textwidth]{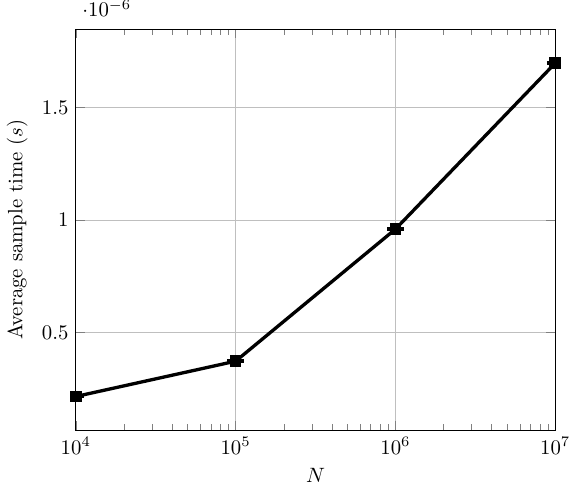}
 \caption{Average sample time from a Complete Binary Tree for different values of the number of events $N$, expressed in seconds. The x-axis is in logarithmic scale. The performance of this structure is sensitive only to the parameter $N$ and it follows $O(\log N)$, although the performance degrades after $\approx 10^7$. Only a rate distribution (the uniform distribution) is shown, as it does not affect performance. The average update time, not shown, has a similar behaviour.}
 \label{fig:CBT-N}
 \end{center}
\end{figure}

We opted for an object-oriented implementation of the Complete Binary Tree, slightly more complex than the heap-based implementation mentioned in Section \ref{sec:tree} but more flexible.
It is clear in Figure~\ref{fig:CBT-N} that the sample time is proportional to $\log(N)$, as we expected from Theorem \ref{thm:tree}. The average update time, not shown, follows the same pattern.

Performance degraded significantly when we tried to push the simulation to values of $N$ larger than shown in Figure~\ref{fig:CBT-N}. One of the underlying assumptions of our work is that we have a vector-like structure that can access a random element in constant time. Once a data structure grows beyond the limits of the cache of the computer we are running our experiment on, we reach slower memory and this assumption is no longer valid.

Different implementations of this method that restrict its memory footprint can in theory allow for larger values of $N$ before hitting the cache memory limits.

\subsection{The Acceptance-Rejection method}
\begin{figure}[htbp]
\begin{center}
 \includegraphics[width=.49\textwidth]{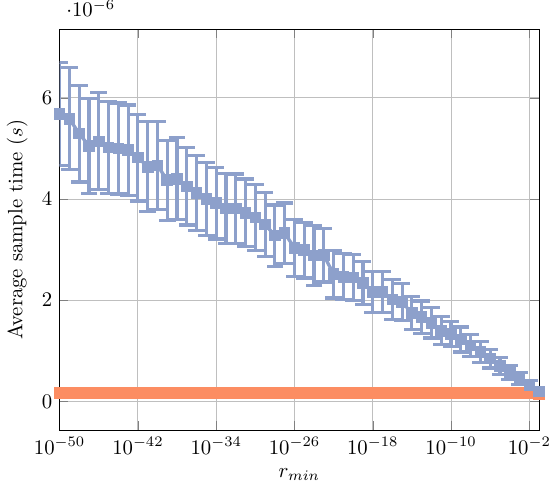}
 \includegraphics[width=.49\textwidth]{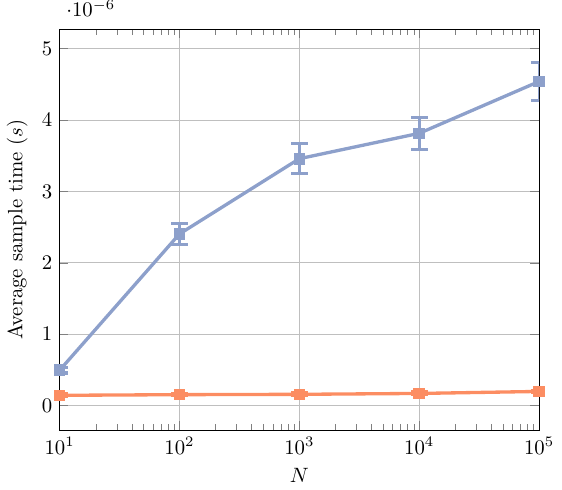}
 \caption{Average sample time from an Acceptance-Rejection structure for different values of the minimum rate $r_{min}$ (left) and the number of events $N$ (right). We show the uniform distribution (orange) and a decreasing distribution with $\beta < 2$  (blue). The sample time clearly follows the expected logarithmic law with the ratio $r_{max}/r_{min}$ for decreasing distributions and it is (relatively) constant for non-decreasing ones. We also note a correlation with $N$ for decreasing distributions: as more events are added to the structure, the discrete probability distribution that we sample from gets closer to the underlying rate distribution, in particular for small values of rate, which affects the performance of the structure.}
 \label{fig:AR-exp}
 \end{center}
\end{figure}

We implemented the Acceptance-Rejection method with a dynamic maximum, i.e. the maximum value is set to the largest value encountered so far, which is clearly $\leq r_{max}$. The Acceptance-Rejection performance appears insensitive to the range of rates for non-decreasing rate distributions, as we can see in Figure~\ref{fig:AR-exp} and expected from Theorem \ref{th:ARNonDecr}. For decreasing rate distributions we notice that the performance degrades linearly with the ratio $r_{max}/r_{min}$, as expected from Lemma \ref{coroll:expAR}. 

We note that it seems to be a correlation with the number of possible events $N$, if they follow a decreasing distribution. As more events are added to the structure, the discrete probability distribution that we sample from gets closer to the underlying rate distribution, in particular for small values of rate, which affects the performance of the structure. Similarly to what we mentioned in the previous section, we also expect further performance degradation for larger values of $N$ once we hit the cache memory limit and we lose constant time access to the vector containing the events.

\subsection{Tree of Groups}
\begin{figure}[htbp]
\begin{center}
 \includegraphics[width=.49\textwidth]{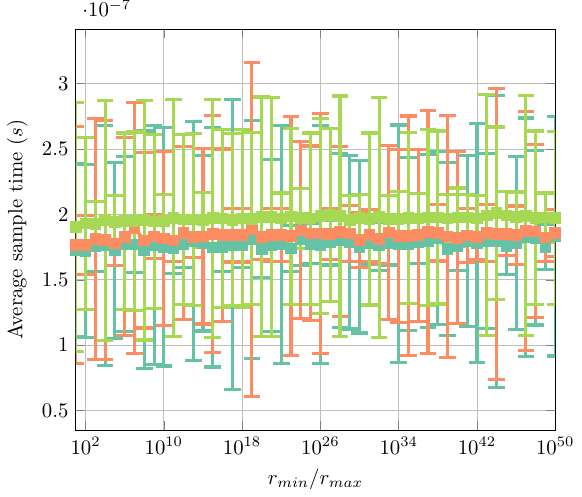}
 \includegraphics[width=.49\textwidth]{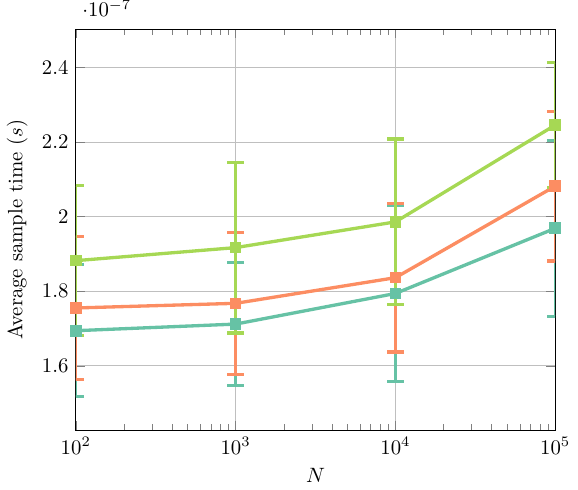}
 \caption{Average sample time from a Tree of Groups data structure for different values of  $r_{max}/r_{min}$ (left) and the number of events $N$ (right). We show the increasing distribution (light blue), uniform distribution (orange) and a decreasing distribution with $\beta > 2$  (green). The expected degradation of performance for this data structure is so slow ($O(\log\log r_{max}/r_{min})$) that it appears constant with regards to $r_{min}$. We note that the sample time for the decreasing distribution grows with $N$ for the previously described statistical effect and both grow for $N=10^5$, which is when we start hitting the cache memory limit in this experiment. }
 \label{fig:tog-exp}
 \end{center}
\end{figure}
The Tree of Groups method has similar performance for all rate distributions and, as we can see in Figure~\ref{fig:tog-exp}, it does not seem to be significantly sensitive even to the range of the rates. The asymptotic behaviour of the sample time, which we expect to be $O\left( \log \log \frac{r_{min}}{r_{max}}\right)$, is too small to be noticeable even for extremely large values of $\frac{r_{max}}{r_{min}}$.

There is a minor correlation with the number of events $N$ for the decreasing distribution that we can explain with the same arguments of the previous subsection. Nevertheless, since the complexity grows with the logarithm of the ratio the statistical effect for small values of $N$ is significantly less dramatic than previously. Finally, the average sample time grows for all rate distributions once we reach $N = 10^5$, which is when we start hitting the cache memory limit in this experiment.

\subsection{Cascade of Groups}
\begin{figure}[!htbp]
\begin{center}
 \includegraphics[width=.7\textwidth]{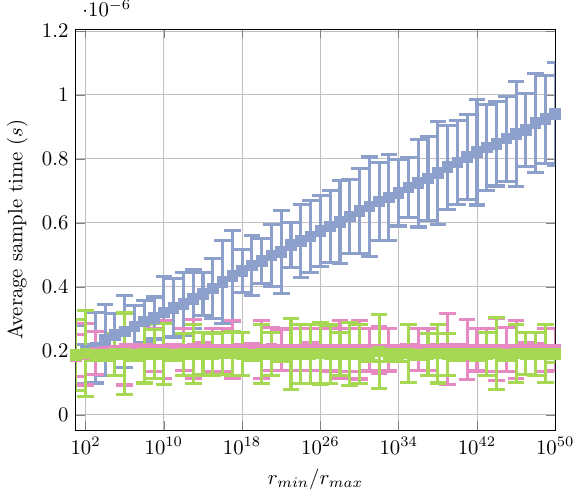}
 \caption{Average sample time from a Cascade of Groups data structure for different values of  $r_{max}/r_{min}$. We show three decreasing distributions: $\beta < 2$ (blue), $\beta = 2$ (pink) and $\beta > 2$ (green). The experimental results confirms the constant time sampling for $\beta \geq 2$, which is a stronger results than our Theorem~\ref{theorem:cascade}. The effect of $N$ on the sampling time is negligible in comparison to the difference between rate distributions. }
 \label{fig:cog-exp}
 \end{center}
\end{figure}
The Cascade of Group shows a stronger result we expected from Theorem~\ref{theorem:cascade}. As we can see in Figure~\ref{fig:cog-exp} it guarantees expected constant time sampling for rate distributions that decrease at least as fast as $1/r^2$. The effect of $N$ on the sampling time is negligible in comparison to the difference between rate distributions. 

\subsection{Reverse Cascade of Groups}
\begin{figure}[!htbp]
\begin{center}
 \includegraphics[width=.7\textwidth]{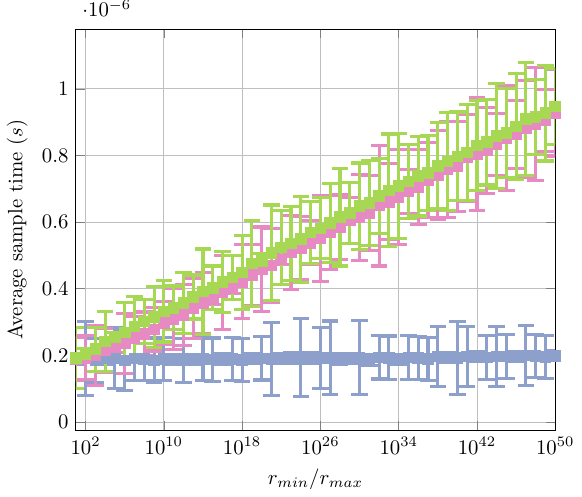}
 \caption{Average sample time from a Reverse Cascade of Groups data structure for different values of  $r_{max}/r_{min}$. We show three decreasing distributions: $\beta < 2$ (blue), $\beta = 2$ (pink) and $\beta > 2$ (green). The result is, as expected, exactly the opposite of the Cascade of Groups: the sampling is performed in expected constant time for $\beta < 2$. The effect of $N$ on the sampling time is negligible in comparison to the difference between rate distributions. }
 \label{fig:revcog-exp}
 \end{center}
\end{figure}
Quite appropriately, the Reverse Cascade of Groups has the opposite result of the Cascade of Group; as we can see in Figure~\ref{fig:revcog-exp}, the structure performs the sampling in expected constant time for rate distributions that decrease slower than $1/r^2$, which is in line with our Theorem~\ref{theorem:reversecascade}. The effect of $N$ on the sampling time is negligible in comparison to the difference between rate distributions. 

\subsection{Two Levels  Acceptance-Rejection}
\begin{figure}[htbp]
\begin{center}
 \includegraphics[width=.49\textwidth]{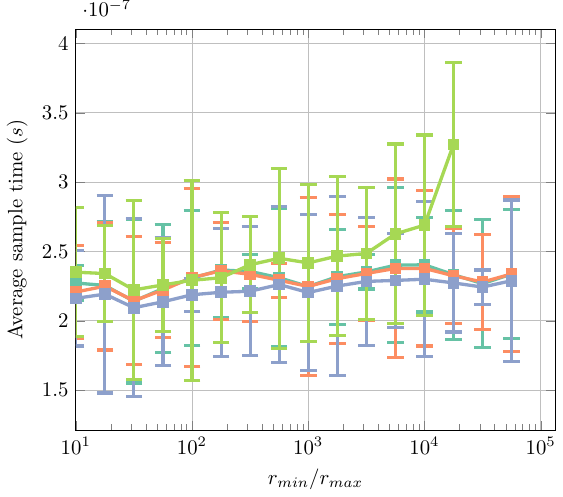}
 \includegraphics[width=.49\textwidth]{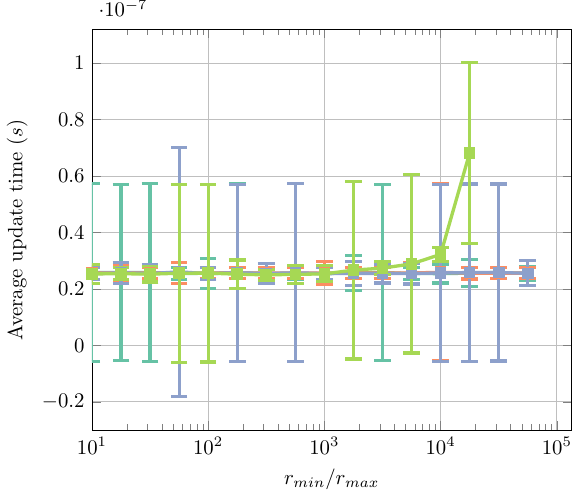}
 \caption{Average sample (left) and update (right) time from a Two Levels  Acceptance-Rejection data structure for different values of  $r_{max}/r_{min}$, constructed such that $R_{tot} > B \; g$. We show four distributions: an increasing distribution (light blue), a uniform distribution (orange), a decreasing distribution with $\beta < 2$  (blue) and one with $\beta > 2$ (green). Under its assumption, this method guarantees both constant time update and sampling, but we note that the number of events required to satisfy the condition grows with both the ratio $r_{max}/r_{min}$ and faster decreasing distributions, to the point where the cache memory limit is clearly hit at around $r_{max}/r_{min} = 10^4$ for $\beta > 2$.}
 \label{fig:2arb-exp}
 \end{center}
\end{figure}
We implemented the Two Levels  Acceptance-Rejection that we previously described, setting $B = r_{max}$ and $c = 2$. Theorem~\ref{th:2AR-MB} tells us that the performance of this data structure is correlated with the amount of total rate inside it: if $R_{tot} > B \; g$, both sampling and update should require constant time, regardless of the rate distribution. Events are therefore added to the data structure until this condition is satisfied, implying that $N$ is no longer an experimental parameter in our control. We also note that the number of events required for this condition becomes rapidly large for (faster) decreasing distributions and bigger ratio ranges. 

As we can see in Figure~\ref{fig:2arb-exp}, this data structure does in fact guarantee constant time samples and updates, albeit with some variability, as long as the condition on the amount of rate is satisfied. Unfortunately, the number of events required for decreasing distributions quickly fill the cache memory and already for $r_{max}/r_{min} = 10^4$, in our specific implementation and system, we lose the expected constant time performance.

\section{Conclusions}
\label{section:conclusions}
In this work, we have presented two basic data structures for sampling from a discrete probability distribution, the Acceptance-Rejection method and the Complete Binary Tree, and used them as building blocks for some multi-level data structures for the dynamic case: the Tree of Groups, the Cascade of Groups and Two Levels Acceptance-Rejection. 

We have proved, under our assumptions, constant time sampling and updates for different classes of rate distributions and a generic result that requires an assumption on the amount of rate in the structure. These results have been confirmed by our experimental analysis, which has also highlighted the practical advantages of the Tree of Groups when faced with real-life constraints and the downsides of the theoretically optimal Two Levels  Acceptance-Rejection.  Multilevel methods allowed us both to optimize the sampling to the particular conditions of the problem, and obtain significant general results. 

While inspired by a practical application, our set of assumptions is arbitrary. Further study is warranted for other sets of assumptions, both inspired by theoretical interest and realistic applications. For instance, we could make assumptions on the updates and assume that the rates are increased or decreased by a known constant quantity when updated while removing other assumptions. 


\newpage
\bibliographystyle{acm}
\bibliography{biblio}
\end{document}